\documentstyle[12pt]{article}
\newcommand{\beq}{\begin{equation}}
\newcommand{\eeq}{\end{equation}}

\newcommand{\eq}[1]{eq.(\ref{#1})}
\title {
\begin{flushright} PSU/TH/129
\end{flushright}
\bigskip
FIRST CORRECTIONS TO HYPERFINE SPLITTING AND LAMB SHIFT
INDUCED BY DIAGRAMS WITH TWO EXTERNAL PHOTONS AND SECOND ORDER RADIATIVE
INSERTIONS IN THE ELECTRON LINE} \medskip

\author {Michael I. Eides\\
Department of Physics, Pennsylvania State University,\\
University Park, Pennsylvania 16802, USA \thanks{E-mail address:
eides@phys.psu.edu}\thanks{Temporary address up to August 31, 1993}\\
and\\
Petersburg Nuclear Physics Institute,\\
Gatchina, St.Petersburg 188350, Russia\thanks{E-mail address:
eides@lnpi.spb.su}
\thanks{Permanent address}
\\
and\medskip\\
\and
Savely G. Karshenboim\hspace{.5cm} and \hspace{.5cm} Valery A.
Shelyuto\medskip
\\ D. I.  Mendeleev Institute of Metrology, \\ St.Petersburg
198005, Russia}
\date{June 10, 1993}

\begin{document}
\baselineskip-15cm
\maketitle
\lineskip3pt
\thispagestyle{empty}

\newpage
\thispagestyle{empty}
\begin{abstract}
Contributions to HFS and to the Lamb shift intervals of order
$\alpha^2(Z\alpha)^5$ induced by graphs with two radiative photons
inserted in the electron line are considered. It is demonstrated that
this last gauge invariant set of diagrams which are capable of
producing corrections of considered order consists of nineteen topologically
different diagrams. Contributions both to HFS and Lamb shift induced by
graphs containing one-loop electron self-energy as a subgraph and also by
the graph containing two one-loop vertices are obtained.
\end{abstract}

\newpage
\setcounter{page}{1}
\section {Introduction}
Recent theoretical work on high order corrections to hyperfine splitting
and Lamb shift concentrated on calculation of nonrecoil contributions of
order $\alpha^2(Z\alpha)^5$. These corrections are of immediate
phenomenological interest for HFS measurements in the ground state of
muonium and for $n=2$ Lamb shift measurements in hydrogen. Their magnitude
may run up to several kilohertz for both intervals to be compared with
current experimental uncertainty equal to $0.16 kHz$ for HFS in muonium
\cite{Mariam} and to $1.9 kHz$ for the Lamb shift in hydrogen
\cite{Sokolov}. Further experimental progress is envisaged at least in the
case of muonium hyperfine splitting measurements \cite{Hughes}.

The only terms, which are capable to produce corrections  with magnitude
about several kilohertz to muonium HFS (an order of magnitude larger than
the experimental error) and are still uncalculated are of functional form
$\alpha^2(Z\alpha)E_F$ (see, e.g. \cite{ann1}, \cite{ann2}\footnote{Several
leading logarithmic corrections of higher
order which nevertheless turn out to be numerically significant were
recently obtained in \cite{k}.}). For the Lamb shift in hydrogen corrections
of order $\alpha^2(Z\alpha)^5m$ are also as large as several kilohertz but
in this case there are also other sources of contributions of comparable
magnitude (see, e.g.  \cite{ego}).

We have shown recently that there  exist  six  gauge  invariant
sets of diagrams (see Fig.1), which produce corrections of order
$\alpha^2(Z\alpha)^5$ to muonium HFS \cite{eks1}. All these diagrams may
be obtained with the help of different radiative insertions from the
skeleton diagram, which contains two external photons attached to the
electron line.  Contributions induced by polarization operator  insertions
in external photons and by simultaneous insertion of radiative photon in
electron line and one-loop  polarization operator in external photon had
been calculated  in analytic form \cite{eks1}. Correction induced by
polarization operator insertions  in  radiative  photons  was obtained in
semianalytic form as one-dimensional integral, where  integrand  is
itself  a complete elliptic integral \cite{eks2}. Contribution induced by
the fifth gauge invariant set of graphs containing light by light
scattering insertions was reduced to the three dimensional integral over
Feynman parameters which was calculated numerically \cite{eks5}.

Contributions to  the Lamb shift of order $\alpha^2(Z\alpha)^5$ are
induced by the same six gauge invariant sets of diagrams which were just
described for hyperfine structure. The only difference is that the  tensor
structure which is relevant for the Lamb shift differs from the one which
was relevant for HFS.  Numerical calculation of contributions induced by
diagrams with polarization operator insertions in the external photons
and in the radiative photons as well as by the graphs containing light by
light scattering insertions was performed recently \cite{ego}, \cite{eg}
\cite{eg4} and \cite{eg5}.

In the present paper we discuss all possible contributions  to  HFS and
to the Lamb shift of order $\alpha^{2}(Z\alpha)^{5}$ which are
induced by the last and most bulky set of diagrams with two radiative
photons inserted in the electron line (see representative graph in Fig.1c).
We present below results of calculation of all contributions both to HFS
and to the Lamb shift induced by diagrams containing one-loop electron
self-energy as a subgraph and also by the diagram containing two
one-loop vertices\footnote {Prof. T. Kinoshita announced recently in the
talk presented at the International Workshop on Low Energy Muon Science,
Santa Fe, NM, April 4 - 8, 1993 preliminary result of his calculation of the
contribution to HFS of order $\alpha^2(Z\alpha)^5$ induced by graphs in
Fig.1c (Preprint CLNS 93/1219, May 1993).}'\footnote{Dr. K. Pachucki is
also working now on calculation of contribution of order
$\alpha^2(Z\alpha)^5$ induced by graphs in Fig.1c to the Lamb shift.
(Private communication from  K. Pachucki to M. Eides).}.

\section {General Strategy of Calculations}

The set of diagrams with insertions of two radiative photons in the
electron line which is relevant for calculation of corrections of
order $\alpha^2(Z\alpha)^5$ to HFS and the Lamb shift contains nineteen
topologically different diagrams and is presented in Fig.2. The simplest
way to describe these graphs is to realize that they were obtained from
three graphs for the two-loop electron self-energy by insertion of two
external photons in all possible ways. Really, graphs $2a-2c$ are obtained
from the two-loop reducible electron self-energy diagram, graphs $2d-2k$
are the result of all possible insertions of two external photons in the
rainbow self-energy diagram, and diagrams $2l-2s$ are connected with the
overlapping two-loop self-energy graph.

As was discussed at length in \cite{ann1} (see also \cite{ego} for the case
of the Lamb shift) one has to perform a rather tedious analysis to find out
which diagrams and in what kinematic conditions are relevant for
calculation of the energy shifts. Happily, to this end we may use the
results of just mentioned papers and we know that to obtain contribution to
the energy splitting one has to calculate matrix elements of the diagrams
in Fig.2 between free electron spinors with all external electron lines on
mass-shell, project these matrix elements on the respective spin state and
multiply the result by the square in the origin of the
Schr\"{o}dinger-Coulomb wave function.

It should be mentioned that some diagrams in Fig.2 contain
also contributions of previous order in $Z\alpha$. Physical nature of these
contributions is especially transparent in the case of HFS. As was mentioned
in \cite{sty} they correspond to anomalous magnetic moment, their true
order in $Z\alpha$ is lower than their apparent order and they should be
subtracted from the matrix elements prior to calculation of the
contributions to HFS. Analogous situation holds also in the case of the
Lamb shift. The only difference is that this time not only the slope of the
Pauli formfactor but also the slope of the Dirac formfactor of the electron
is capable to produce lower order contribution to the splitting of the
energy levels (see, e. g. \cite{ego}, \cite{eg} and \cite{eg4}) and all
respective terms have to be subtracted. Technically cases of lower order
contributions both to HFS and to the Lamb shift are quite similar. Lower
order terms are each time proportional to the exchanged momentum squared
and to get rid of them one has to subtract all low frequency terms
proportional to the exchanged momentum squared in the asymptotes of the
matrix elements when they exist.

Actual calculation of matrix elements of the diagrams in Fig.2 is impeded
by the ultraviolet (UV) and infrared (IR) divergencies. To get rid of UV
problems we work only with renormalized (subtracted) graphs and subgraphs
and only such graphs are presented in Fig.2. This is the reason why some
obvious diagrams with e.g. self-energies on the external electron lines are
absent in Fig.2. IR problems are as usual more difficult to deal with than
the UV ones. We choose the Fried-Yennie (FY) gauge for the radiative photons.
The advantage of this choice of gauge is connected with maximally mild
behavior of all matrix elements in the IR region of integration momenta
in the FY gauge and with the possibility to perform on-mass-shell
renormalization without introduction of the infrared photon mass (see,e.g.
\cite{ann1}).  This last feature of the FY  gauge is especially
important for us as we want to push analytical calculations as far forward
as possible and working without the photon mass gives us the chance to
obtain much simpler analytical formulae. On the other hand working in the
FY gauge without the photon mass makes it absolutely necessary to
pay special attention to the infrared behavior of the integrand functions
and to perform cancellation of spurious IR divergences prior to
integration.

Each contribution of order $\alpha^{2}(Z\alpha)^{5}$ arises from
radiative insertions in the skeleton graph in Fig.3 with two external
photons. Contribution to hyperfine splitting, produced by  diagrams in
Fig.2 is given  by  the  expression  (see, e.g. \cite{ann1})

\beq                                  \label{hfs}
\Delta E_{HFS} = 8 \frac{Z\alpha}{\pi n^3}(\frac{\alpha}{\pi})^2\:E_F
\int_0^\infty\frac{d|{\bf k}|}{{\bf k}^2}\:F({\bf k})\:,
\eeq

where $k_\mu=(0,{\bf k})$ is the momentum of the external Coulomb photons
measured in the electron mass units\footnote{All momenta below are measured
in the electron mass units if reverse is not stated explicitly} and

\beq
E_F=\frac{8}{3}(Z\alpha)^4 \frac{{m_r}^3}{mM}(1+a_\mu),
\eeq

where $m_r={m}/(1+{m}/{M})$ is the reduced mass of the electron-muon system
and $a_\mu$ is the anomaly of the muon magnetic moment. The function $F({\bf
k})$ is connected with the numerator structure of  each  particular graph
and describes radiative corrections to the skeleton diagram. It is
normalized on the skeleton numerator contribution.

As was already discussed above anomalous magnetic moment of the electron
which is also produced by the radiative insertions in the electron line in
the graphs in Fig.2 is not sufficiently mild near the lower boundary of
integration region over exchanged momentum in eq.(\ref{hfs}) and naively
leads to divergence of the integral for the energy splitting.  This simply
means that anomalous magnetic moment produces contribution of lower order
in $Z\alpha$ and respective terms should be subtracted from the expression
for the electron factor $F({\bf k})$ prior to integration.

Contribution to the Lamb shift induced by the diagrams in Fig.1 has the
form (see, e.g. \cite{ego})

\beq                                \label{lamb}
\Delta E_L =
-\frac{16}{n^3}\:m\:\left(\frac{m_r}{m}\right)^3
\left(\frac{\alpha}{\pi}\right)^2
\frac{(Z\alpha)^5}{\pi }
\int_0^\infty\frac{d|{\bf k}|}{{\bf k}^4}\:L({\bf k})\:.
\eeq

Function $L({\bf k})$ as in the case of the contribution to HFS is
calculated for each particular graph and describes radiative corrections
to the skeleton diagram. Spin structure which is relevant for the Lamb
shift differs, of course, from the one relevant for the HFS. The integral
for the Lamb shift interval in eq.(\ref{lamb}) is more singular in the
infrared region than the one for the HFS in eq.(\ref{hfs}). This
singularity once again indicates that some diagrams in Fig.2  contain not
only contributions of order $\alpha^{2}(Z\alpha)^{5}m$ to the Lamb shift
but also contain abundantly contributions of lower order in  parameter
$Z\alpha$ (compare the contribution of the anomalous magnetic moment of
the electron to HFS discussed above). These contributions are connected
with integration over external photon momenta of characteristic atomic
order $mZ\alpha $ and they should be subtracted to make calculation of
corrections of order $\alpha^{2}(Z\alpha)^{5}m$ possible \cite{ego}.

\section {Calculation of Simplest Contributions}

\subsection {Diagram with Two One-Loop Self-Energy Insertions}
\subsubsection{Contribution to HFS Interval}

We begin actual calculation with the simplest possible graph in Fig.2a,
containing reducible electron self-energy insertion in the skeleton graph.
Explicit expression for the one-loop electron self-energy in the
FY gauge has the form (see, e.g. \cite{ann1})

\beq            \label{sigma}
\Sigma^{FY}(p-k)=(\hat p-\hat k-1)^2(\frac{-3\alpha(\hat{p}-\hat{k})}{4\pi
}) M({\bf k}),
\eeq

where

\beq
M({\bf k})=\frac{1}{1-{\bf k}^2}+\frac{{\bf k}^2}{(1-{\bf k}^2)^2}
\log {\bf k}^2
\eeq

and kinematic conditions are defined by relations
$p_\mu=(1,{\bf 0})$, $pk=0$, $k_\mu=(0,{\bf k})$.

Substitution of this expression for the self-energy operator instead of the
electron factor $F({\bf k})$ in eq.(\ref{hfs}) leads to the one-dimensional
integral for the contribution to the HFS

\beq
\Delta E_{HFS}^{a}=\frac{\alpha^2(Z\alpha)}{\pi
n^3}E_F(-\frac{9}{2\pi^2})\:\int_0^\infty{d|{\bf k}|}
(1-{\bf k}^2)\:M^2({\bf k})\:,
\eeq

which may be easily calculated analytically

\beq
\Delta E_{HFS}^{a}=\frac{9}{4}\frac{\alpha^2(Z\alpha)}{\pi
n^3}E_F.
\eeq

\subsubsection{Contribution to the Lamb Shift}

It is not difficult to see that due to the tensor structure of the
electron-line factor with two one-loop electron self-energy insertions it
does not lead to nonvanishing contribution to the Lamb shift of the order
under consideration

\beq
\Delta E_{L}^{a}=0.
\eeq

\subsection {Diagrams with Simultaneous Insertions of One-Loop Electron
Self-Energy and Vertex}

\subsubsection{General Expression for the One-Loop Vertex with One
On-Mass-Shell Leg in the Fried-Yennie Gauge}

Renormalized vertex operator with on-mass-shell left leg enters as a
subgraph in the diagram in Fig.2b.  We obtained general expression for
those entries in such vertex operator in the FY gauge which
produce contributions to the hyperfine splitting and the Lamb shift (terms
which are proportional to $k_\mu$ are omitted because they are irrelevant
for our goals)

\beq
\label{fyf}
\Lambda_\mu^{FY}=\frac{\alpha}{2\pi}\{A({\bf k}){\bf k}^2\gamma_\mu
+B({\bf k})\gamma_\mu(\hat p-\hat k-1)
\eeq
\[
+C({\bf k})p_\mu(\hat p-\hat
k-1)+E({\bf k})\sigma_{\mu\nu}k^\nu\},
\]

where

\beq
A({\bf k})=-(\frac{2}{|{\bf k}|^3}+\frac{1}{2|{\bf k}|})\Phi({\bf k})
+\frac{2}{{\bf k}^2}S({\bf k})
\eeq
\[
-\frac{3}{2}M({\bf k})-2\frac{\log{{\bf k}^2}}{{\bf k}^2}
-\frac{3}{2}\frac{\log{{\bf k}^2}}{1-{\bf k}^2},
\]

\beq
B({\bf k})=-(\frac{1}{|{\bf k}|}+\frac{|{\bf k}|}{8})\Phi({\bf k})
+\frac{1}{2}S({\bf k})-\frac{5}{4}M({\bf k})
+\frac{1}{4}
\eeq
\[
-\frac{1}{8}\log{{\bf k}^2}-\frac{7}{8}\frac{\log{{\bf k}^2}}{1-{\bf k}^2},
\]

\beq
C({\bf k})=\frac{1}{|{\bf k}|}\Phi({\bf k})-S({\bf k})-\frac{1}{2}M({\bf k})
+\frac{1}{2}\log{{\bf k}^2}-\frac{3}{2}\frac{\log{{\bf k}^2}}{1-{\bf k}^2},
\eeq

\beq
E({\bf k})=-\frac{|{\bf k}|}{8}\Phi({\bf k})-\frac{1}{2}S({\bf k})
-\frac{1}{4}M({\bf k})
\eeq
\[
+\frac{1}{4}+\frac{3}{8}\log{{\bf k}^2}-\frac{3}{8}\frac{\log{{\bf k}^2}}
{1-{\bf k}^2}
\]

and

\beq  \label{phi}
\Phi({\bf k})=|{\bf k}|\int_0^1\frac{dz}{1-{\bf k}^2z^2}\log{\frac{1
+{\bf k}^2z(1-z)}{{\bf k}^2z}}
=\rm{Li}(1-|{\bf k}|)-\rm{Li}(1+|{\bf k}|)
\eeq
\[
+2\left[\rm{Li}(1+\frac{\sqrt{{\bf
k}^2+4} +|{\bf k}|}{2}) -\rm{Li}(1-\frac{\sqrt{{\bf
k}^2+4}+|{\bf k}|}{2})-\frac{\pi^2}{4}\right],
\]
\[
S({\bf
k})=\frac{\sqrt{{\bf k}^2+4}}{2|{\bf k}|}\log{ \frac{\sqrt{{\bf k}^2+4}+|{\bf
k}|}{\sqrt{{\bf k}^2+4}-|{\bf k}|}}.
\]

Euler dilogarithm Li is defined here as in \cite{ann1}. We would like to
mention that function $\Phi({\bf k})$ emerged for the first time in
calculation of contribution to HFS induced by the diagrams in Fig.1c
\cite{eks1}.

\subsubsection{Contribution to HFS Interval}

It is not difficult now to obtain contribution induced by diagrams in
Fig.2b to HFS. Taking into account combinatorial coefficient 2 we obtain

\beq
2\Delta E_{HFS}^{b}=\frac{\alpha^2(Z\alpha)}{\pi
n^3}E_F(-\frac{6}{\pi^2})\:\int_0^\infty{d|{\bf k}|}\:M({\bf k})
\:[-{\bf k}^2A({\bf k})+B({\bf k})+\frac{1}{2}C({\bf k})]
\eeq

or numerically

\beq
2\Delta E_{HFS}^{b}=-6.65997(1)\frac{\alpha^2(Z\alpha)}{\pi
n^3}E_F.
\eeq

\subsubsection{Contribution to the Lamb Shift}

Contribution to the Lamb shift induced by the diagrams in Fig.2b has the
form

\beq
2\Delta E_{L}^{b}
=\:m\:\left(\frac{m_r}{m}\right)^3
\frac{\alpha^2(Z\alpha)^5}{\pi n^3}\left(\frac{6}{\pi^2}\right)
\int_0^\infty{d|{\bf k}|}M({\bf k})[-A({\bf k})+B({\bf k})
\eeq
\[
+C({\bf k})
-E({\bf k})]
\]

or numerically

\beq
2\Delta E_{L}^{b}=2.9551(1)
\:m\:\left(\frac{m_r}{m}\right)^3
\frac{\alpha^2(Z\alpha)^5}{\pi n^3}.
\eeq

\subsection {Diagram with Simultaneous Insertions of Two One-Loop Vertices}

\subsubsection{Contribution to HFS Interval}

Contribution to HFS induced by the diagram in Fig.2c has the form

\beq
\Delta E_{HFS}^{c}=\frac{\alpha^2(Z\alpha)}{\pi
n^3}E_F(\frac{2}{\pi^2})\:\int_0^\infty{d|{\bf k}|}\:\{{\bf k}^2A^2({\bf k})
\eeq
\[
-2A({\bf k})E({\bf k})
-[E({\bf k})-B({\bf k})]^2+C({\bf k})[E({\bf k})-B({\bf k})]\}
\]

or numerically

\beq
\Delta E_{HFS}^{c}=3.93208(1)\frac{\alpha^2(Z\alpha)}{\pi
n^3}E_F.
\eeq

\subsubsection{Contribution to the Lamb Shift}

Contribution to the Lamb shift induced by the diagrams in Fig.2c has the
form

\beq
\Delta E_{L}^{c}
=\:m\:\left(\frac{m_r}{m}\right)^3
\frac{\alpha^2(Z\alpha)^5}{\pi n^3}\left(\frac{4}{\pi^2}\right)
\int_0^\infty{d|{\bf k}|}A({\bf k})[-A({\bf k})+B({\bf k})
\eeq
\[
+C({\bf k})
-E({\bf k})]
\]

or numerically

\beq
\Delta E_{L}^{c}=-2.2231(1)
\:m\:\left(\frac{m_r}{m}\right)^3
\frac{\alpha^2(Z\alpha)^5}{\pi n^3}.
\eeq

\section {Calculation of Contributions Induced by the Diagrams with
Insertions of the "Left" Self-Energy Operator}

\subsection {Diagrams Containing the Two-Loop Vertex}

\subsubsection {General Expression for the "Left" Vertex with Left
One-Loop Self-Energy Insertion}

Let us consider two-loop vertex diagram which is contained as a subgraph in
Fig.2d.  Insertion of the one-loop electron self-energy in the one-loop
vertex  corresponds in the FY gauge to the replacement

\beq        \label{sesubl}
\frac{1}{\hat p+\hat q-1}\rightarrow \frac{3\alpha}{4\pi}(\hat p+\hat q)
\int_0^1dx\frac{x}{1-x}\frac{1}{q^2+2pq-\frac{x}{1-x}}
\eeq

of the left propagator in the one-loop vertex integrand. We used here
one-dimensional integral representation for the renormalized self-energy
operator in the FY gauge and took into account that the would
be loop integration momentum $q$ unlike momentum $k$ of the external
photon in eq.(\ref{sigma}) is not orthogonal to momentum $p$ of the
external electron.

Formal expression for the vertex has the form

\beq                           \label{leftvert}
\Lambda_\mu^{d}
=3(\frac{\alpha}{4\pi})^2 \int_0^1dx\frac{x}{1-x}
\int\frac{d^4{q}}{i\pi^2}\frac{N_F^d+N_L^d/q^2}{q^2[q^2+2q(p-k)+k^2](q^2+2pq
-\frac{x}{1-x})]},
\eeq

where

\beq
N_F^{d}= \gamma^\sigma(\hat p+\hat q)\gamma_\mu(\hat p+\hat q-\hat
k+1)\gamma_\sigma,
\eeq
\beq
N_L^{d}={2}\hat q(\hat p+\hat q)\gamma_\mu(\hat p+\hat q-\hat
k+1)\hat q.
\eeq

It is convenient to put down second numerator in the form

\beq
N_L^{d}=q^2N_{L1}^{d}+N_{L2}^{d},
\eeq

where

\beq
N_{L1}^{d}=2[\gamma_\mu(\hat p-\hat k+1)\hat q+\hat q\hat
p\gamma_\mu+q^2\gamma_\mu],
\eeq
\[
N_{L2}^{d}=2\hat q\hat p\gamma_\mu(\hat p-\hat k+1)\hat q.
\]

As a result of this transformation numerator $N_{L1}^{d}$ enters all
expressions below on the same footing as $N_F^{d}$. Combining denominators
in eq.(\ref{leftvert})

\beq      \label{comb}
(1-t)q^2+t[u(q^2+2q(p-k)+k^2)
\eeq
\[
+(1-u)(q^2+2pq-\frac{x}{1-x})]=(q+Q)^2-\Delta_l,
\]
\[
Q=pt-kut,
\]
\[
\Delta_l=t[{\bf k}^2u(1-ut)+t+\frac{x(1-u)}{1-x}]\equiv ta_l,
\]

we obtain after shift of the integration variable $q\rightarrow
q-Q$ representation for the bare vertex operator

\beq
\Lambda_\mu^{d}=3(\frac{\alpha}{4\pi})^2 \int_0^1dx\frac{x}{1-x}
\int_0^1dt\int_0^1du\int\frac{d^4q}{i\pi^2}
\eeq
\[
\{2t\frac{N_F^{d}(q-Q)
+N_{L1}^{d}(q-Q)}{(q^2-\Delta_l)^3}
\]
\[
+6t(1-t)\frac{N_{L2}^{d}(q-Q)}{(q^2-\Delta_l)^4}\}.
\]

Momentum integration leads to

\beq                                 \label{unsubtr}
\Lambda_\mu^{d}=3(\frac{\alpha}{4\pi})^2 \int_0^1dx\frac{x}{1-x}
\int_0^1dt\int_0^1du\{[6t(\log{\frac{\Lambda^2+\Delta_l}{\Delta_l}}
\eeq
\[
-\frac{3}{2}\frac{\Lambda^2}{\Lambda^2+\Delta_l})
-t\frac{4{\bf k}^2ut(1-ut)+2(1-3t)}{\Delta_l}]\gamma_\mu
+\frac{t}{\Delta_l}2t^2\gamma_\mu(\hat p-\hat k-1)
\]
\[
+\frac{4t}{\Delta_l}(1-2t+t^2u)p_\mu(\hat p-\hat k-1)
+\frac{2t}{\Delta_l}[(1-t)^2-ut]\gamma_\mu\hat k
\]
\[
+2(1-t)[-\frac{t}{\Delta_l}+\frac{t^2}{\Delta_l^2}(-2{\bf
k}^2ut+2t)]\gamma_\mu
\]
\[
+\frac{t^2}{\Delta_l^2}2t(1-t)[-{\bf
k}^2u(1-u)+1]\gamma_\mu(\hat p-\hat k-1)
\]
\[
+4(1-t)[\frac{t}{\Delta_l}-\frac{t^2}{\Delta_l^2}t]p_\mu(\hat
p-\hat k-1)
\]
\[
+2(1-t)[\frac{t}{\Delta_l}-\frac{t^2}{\Delta_l^2}ut({\bf
k}^2(1-2u)+2)]\gamma_\mu\hat k\}.
\]

Subtraction term is equal to

\beq          \label{subtr}
\Lambda_\mu^{d}(0)=3(\frac{\alpha}{4\pi})^2
\int_0^1dx\frac{x}{1-x}
\int_0^1dt\int_0^1du\{[6t(\log{\frac{\Lambda^2+\Delta_{l0}}{\Delta_{l0}}})
\eeq
\[
-\frac{3}{2}\frac{\Lambda^2}{\Lambda^2+\Delta_l})
-\frac{t}{\Delta_{l0}}2(1-3t)
+2(1-t)[-\frac{t}{\Delta_{l0}}+\frac{t^2}{\Delta_{l0}^2}2t]\}\gamma_\mu,
\]

where

\[
\Delta_{l0}\equiv ta_{l0}=\Delta_{l{|{\bf k}}=0}.
\]

Note that we preserved term $\Delta$ when it is added to the ultraviolet
cutoff $\Lambda^2$ in \eq{unsubtr} and \eq{subtr} in the terms which do not
vanish when the cutoff goes to infinity. This was necessary to preserve
convergence of the subsequent integration over the Feynman parameter $x$ in
\eq{unsubtr} and \eq{subtr} in spite of the singular nature of the explicit
integration weight $x/(1-x)$ near $x=1$. Really, it is easy to check that
$\Delta$ itself is singular near $x=1$

\beq
\Delta_{l{|x\rightarrow 1}}\simeq {\Delta_{l0}}_{|x\rightarrow 1} \simeq
\frac{t(1-u)}{1-x}
\eeq

and compensates would be divergence. All potential divergences in
integration over $x$ vanish after subtraction and subtracted vertex admits
the limit of infinite cutoff. All extra terms containing ultraviolet cutoff
$\Lambda^2$  which were preserved in \eq{unsubtr} and \eq{subtr} to
make the integration over $x$ finite vanish in the subtracted expression
for the vertex at least as $(1/\Lambda^2)\log(\Lambda^2)$. Hence, subtracted
expression for the vertex coincides with the naive one which would be
obtained if one simply missed the problem of finiteness of integration over
$x$ in \eq{unsubtr} and \eq{subtr}!

Last terms in \eq{unsubtr} and in \eq{subtr} may be simplified with the
help of identity in \eq{refer}, where one has to substitute $n=1$, $\eta=1$
and $\tau=u$. We also simplify logarithmic term in \eq{unsubtr} and
\eq{subtr} which  admits (as was just explained) omission of the ultraviolet
cutoff after subtraction. We perform integration by parts over $u$,
under which subtracted logarithmic term transforms as

\beq                  \label{2term}
6t\log{\frac{a_{l0}}{a_l}}\rightarrow-6t(1-u)[\frac{{\bf
k}^2(1-2ut)}{a_l}+\frac{x}{1-x}(\frac{1}{a_{l0}} -\frac{1}{a_l})]
\eeq
\[
=6t\frac{{\bf k}^2}{a_l}[1-2ut-u^2t+\frac{ut(1-ut)}{a_{l0}}].
\]

Sum of the expressions in eqs.(\ref{subtr}) and (\ref{2term}) and other
terms in eq.(\ref{unsubtr}) proportional to $\gamma_\mu$ is equal to

\beq                        \label{aa}
{\cal A}_l=-\frac{t}{a_l}(6+4u-12ut+2u^2t)
\eeq
\[
+\frac{6t^2u(1-ut)}{a_la_{l0}}-\frac{4t(1-t)u(1-u)}{a_l^2}.
\]

We also introduce several additional functions in such way as to make final
expression look like the one for the one-loop vertex in eq.(\ref{fyf})

\beq             \label{bcd}
{\cal B}_l=2t[\frac{t}{a_l}+\frac{1-t}{a_l^2}(1-{\bf k}^2u(1-u))],
\eeq
\[
{\cal C}_l=4[\frac{2-3t+t^2u}{a_l}-\frac{t(1-t)}{a_l^2}],
\]
\[
{\cal E}_l=2\frac{2-3t+t^2-ut}{a_l}-\frac{2(1-t)ut}{a_l^2}[{\bf
k}^2(1-2u)+2].
\]

Then renormalized vertex with self-energy insertion has the form

\beq   \label{leftverrgtse}
\Lambda_\mu^{d}=3(\frac{\alpha}{4\pi})^2 \int_0^1dx\frac{x}{1-x}
\int_0^1dt\int_0^1du\{
{\bf k}^2{\cal A}_l\gamma_\mu
\eeq
\[
+{\cal B}_lm\gamma_\mu(\hat p-\hat k-1) +{\cal C}_lp_\mu(\hat p-\hat k-1)
+{\cal E}_l\gamma_\mu\hat k\}.
\]

Note that all four functions ${\cal A}_l({\bf k})$, ${\cal B}_l({\bf k})$,
${\cal C}_l({\bf k})$ and ${\cal E}_l({\bf k})$ are finite at ${\bf k}=0$.

\subsubsection{Contribution to HFS Interval}

It is not difficult now to obtain contribution to HFS

\beq               \label{lefthfs}
2\Delta E_{HFS}^{d}=\frac{\alpha^2(Z\alpha)}{\pi
n^3}E_F(\frac{3}{\pi^2})\int_0^1dx\frac{x}{1-x}
\int_0^1dt\int_0^1du\int_0^\infty d|{\bf k}|[{\cal A}_l({\bf k})
\eeq
\[
-\frac{{\cal E}_l({\bf k})-{\cal E}_l(0)}{{\bf k}^2}],
\]

where

\beq
{\cal A}_l({\bf k})-\frac{{\cal E}_l({\bf k})-{\cal E}_l(0)}{{\bf k}^2}
=-\frac{2t}{a_l}(3+2u-6ut+u^2t)-\frac{2ut(1-t)}{a_l^2}
\eeq
\[
+\frac{2u(1-ut)}{a_la_{l0}}(2-3t+4t^2-ut)-4u^2t(1-t)(1-ut)
(\frac{1}{a_l^2a_{l0}}+\frac{1}{a_la_{l0}^2}).
\]

Subtraction of ${\cal E}_l(0)$ in eq.(\ref{lefthfs}) corresponds to
subtraction of the contribution to HFS induced by the second order
anomalous magnetic moment which is of lower order in parameter $Z\alpha$.

After momentum integration we obtain three-dimensional integral for the
contribution to HFS induced by both diagrams with nonsymmetric self-energy
insertions in Fig.2d

\beq
2\Delta E_{HFS}^{d}=\frac{\alpha^2(Z\alpha)}{\pi
n^3}E_F(\frac{3}{2\pi})\int_0^1dx\frac{x}{1-x}
\int_0^1\frac{du}{\sqrt{u}}\int_0^1\frac{dt}{\sqrt{1-ut}}
\eeq
\[
\{\frac{{f}^{d}_{1/2}}{{a_{l0}}^\frac{1}{2}}+\frac{{f}^{d}_{3/2}}
{{a_{l0}}^\frac{3}{2}}+\frac{f^{d}_{5/2}}{{a_{l0}}^\frac{5}{2}}\},
\]

where

\beq
{f}^{d}_{1/2}=-2t(3+2u-6ut+u^2t),
\eeq
\[
{f}^{d}_{3/2}=2u(1-ut)(2-3t+4t^2-ut)-ut(1-t),
\]
\[
{f}^{d}_{5/2}=-6u^2t(1-t)(1-ut).
\]

Numerically we obtain

\beq
2\Delta E_{HFS}^{d}=-3.903(1)\frac{\alpha^2(Z\alpha)}{\pi
n^3}E_F.
\eeq

\subsubsection{Contribution to the Lamb Shift}

Contribution to the Lamb shift has the form

\beq               \label{leftlmb}
2\Delta E_{L}^{d}=-m\:(\frac{m_r}{m})^3
\frac{\alpha^2(Z\alpha)^5}{\pi n^3}(\frac{6}{\pi^2})\int_0^1dx\frac{x}{1-x}
\eeq
\[
\int_0^1dt\int_0^1du
\int_0^\infty d|{\bf k}|\frac{{\cal L}^{d}({\bf k})-{\cal L}^{d}(0)}
{{\bf k}^2},
\]

where

\beq
{\cal L}^{d}({\bf k})\equiv {\cal A}_l({\bf k})-\frac{1}{2}{\cal B}_l({\bf
k}) -\frac{1}{2}{\cal C}_l({\bf k})+\frac{1}{2}{\cal E}_l({\bf k})
\eeq

and

\beq
\frac{{\cal L}^{d}({\bf k})-{\cal L}^{d}(0)}{{\bf k}^2}
=\frac{t(1-t)u^2}{a_l^2}
+ \frac{u(1-ut)}{a_la_{l0}}[2+3t(1-4t)+ut(5+14t-10ut)]
\eeq
\[
-\frac{6u^2(1-ut)^2}{a_la_{l0}^2}+t(1-t)u(1-ut)(-1+6u-4u^2)
(\frac{1}{a_l^2a_{l0}}+\frac{1}{a_la_{l0}^2}).
\]

Subtraction in eq.(\ref{leftlmb}) corresponds to subtraction of the
contributions to the Lamb shift induced by the slope of the Dirac
formfactor and by the anomalous magnetic moment which are of lower order in
parameter $Z\alpha$.

After momentum integration we obtain three-dimensional integral for the
contribution to the Lamb shift induced by both diagrams with nonsymmetric
self-energy insertions in Fig.2d

\beq
2\Delta E_{L}^{d}=m\:(\frac{m_r}{m})^3
\frac{\alpha^2(Z\alpha)^5}{\pi
n^3}(\frac{3}{2\pi})\int_0^1dx\frac{x}{1-x}
\int_0^1\frac{du}{\sqrt{u}}\int_0^1\frac{dt}{\sqrt{1-ut}}
\eeq
\[
\{\frac{g^{d}_{3/2}}{{a_{l0}}^\frac{3}{2}}
+\frac{g^{d}_{5/2}}{{a_{l0}}^\frac{5}{2}}\},
\]

where

\beq
g^{d}_{3/2}=-2u(1-ut)(2+3t+5ut-10ut^2+2u^2t^2)-t(1-t)u^2,
\eeq
\[
g^{d}_{5/2}=3ut(1-ut)[4ut(1-ut)+(1-t)(1-6u+4u^2)].
\]

Numerically we obtain

\beq
2\Delta E_{L}^{d}=-5.235(2)m\:(\frac{m_r}{m})^3
\frac{\alpha^2(Z\alpha)^5}{\pi n^3}.
\eeq

\subsection {Spanning Photon Diagram with the "Left"
Self-Energy Insertion}

\subsubsection {Contribution to HFS Interval}

Consider now spanning photon  diagram with "left" self-energy insertion in
Fig.2e. Unlike discussion of the vertex in the previous subsection we will
consider here only the part of the spanning photon diagram relevant for
hyperfine splitting. Insertion of the one-loop electron self-energy in the
FY gauge corresponds to replacement described in eq.(\ref{sesubl}) in the
left propagator in the one-loop vertex integrand.

Formal expression for the spanning photon diagram has the form

\beq
\Xi_{\mu\nu}^{e}         \label{genspan}
=3(\frac{\alpha}{4\pi})^2 \int_0^1dx\frac{x}{1-x}
\eeq
\[
\int\frac{d^4{q}}{i\pi^2}\frac{N_F^e+N_L^e/q^2}{q^2[q^2+2q(p-k)+k^2](q^2+2pq
-\frac{x}{1-x})(q^2+2pq)},
\]

where

\beq
N_F^{e}= \gamma^\sigma(\hat p+\hat q)\gamma_\mu(\hat p+\hat q-\hat
k+1)\gamma_\nu(\hat p+\hat q+1)\gamma_\sigma,
\eeq

\beq
N_L^{e}={2}\hat q(\hat p+\hat q)\gamma_\mu(\hat p+\hat q-\hat
k+1)\gamma_\nu(\hat p +\hat q+1)\hat q.
\eeq

It is convenient prior to following transformations to rewrite numerator of
the longitudinal part in the form

\beq          \label{lonnum}
N_L^{e}={2}(q^2+2pq)[q^2\gamma_\mu(\hat p+\hat
q-\hat k+1)\gamma_\nu+\hat q\hat p\gamma_\mu(\hat p+\hat
q-\hat k+1)\gamma_\nu]
\eeq
\[
\equiv (q^2+2pq)[q^2N_{L1}^{e}+N_{L2}^{e}].
\]

Hence, denominator factor $(q^2+2pq)$ in \eq{genspan} cancels with
the same explicit factor in  the longitudinal numerator in
\eq{lonnum} and  we use the identity

\beq                \label{savel}
\int_0^1dx\frac{x}{1-x}\frac{1}{(q^2+2pq-\frac{x}{1-x})(q^2+2pq)}
\eeq
\[
=\int_0^1\frac{dx}{1-x}\frac{1}{(q^2+2pq-\frac{x}{1-x})^2}
\]

to get rid of the same factor in the term containing numerator $N_F^{e}$.

Next we combine denominators as in eq.(\ref{comb})  and after the shift of
integration variable $q\rightarrow q-Q$ (see \eq{comb}) obtain
retaining only those numerator structures which are relevant for
contribution to HFS

\beq  \label{leftsigmsp}
\Xi_{\mu\nu}^{e}=\frac{3}{8}(\frac{\alpha}{\pi})^2 \int_0^1dx\frac{1}{1-x}
\int_0^1du\int_0^1dt
\eeq
\[
\int\frac{d^4{q}}{i\pi^2}\{3(1-u)t^2\frac{N_F^{e}(q-Q)}
{(q^2-\Delta_l)^4}
\]
\[
+\frac{txN_{L1}^{e}(q-Q)}{(q^2-\Delta_l)^3}
+\frac{3t(1-t)xN_{L2}^{e}(q-Q)}{(q^2-\Delta_l)^4}\}.
\]

Next we project on the spinor structure relevant for HFS splittting
(see, e.g. \cite{ekscur} for explicit expression for the respective
projector) and obtain after momentum integration scalar electron factor

\beq        \label{spanleft}
F^{e}=\frac{3}{8}{\bf k}^2 \int_0^1dx\frac{1}{1-x}\int_0^1du\int_0^1dt
\{(1-u)\frac{t(1-3ut)}{a_l}
\eeq
\[
+(1-u)\frac{{\bf k}^2u^2t^2(1-ut)+[-1+t(2-t)(1-ut)]}{a_l^2}
\]
\[
+\frac{(1-ut)x}{a_l}+\frac{(1-t)[1+u(1-t)]x}{a_l^2}\}.
\]

Substituting electron factor in \eq{spanleft} in \eq{hfs} one obtains
contribution to HFS interval

\beq               \label{leftsphfs}
2\Delta E_{HFS}^{e}=16\frac{Z\alpha}{\pi
n^3}(\frac{\alpha}{\pi})^2E_F
\int_0^\infty\frac{d|{\bf k}|}{{\bf k}^2}F^{e}({\bf k})
\eeq
\[
=\frac{\alpha^2(Z\alpha)}{\pi
n^3}E_F(\frac{3}{2\pi})\int_0^1\frac{dx}{1-x}
\int_0^1\frac{du}{\sqrt{u}}\int_0^1\frac{dt}{\sqrt{1-ut}}
\{\frac{f^{e}_{1/2}}{{a_{l0}}^\frac{1}{2}}
+\frac{f^{e}_{3/2}}{{a_{l0}}^\frac{3}{2}}\},
\]

where

\beq
f^{e}_{1/2}=(1-u)t[-3+5(1-ut)]+2x(1-ut)
\eeq

and

\[
f^{e}_{3/2}=(1-u)[-1+t(2-t)(1-ut)]+x(1-t)[1+u(1-t)].
\]

Numerically we obtain

\beq
2\Delta E_{HFS}^{e}=4.566(2)\frac{\alpha^2(Z\alpha)}{\pi n^3}E_F.
\eeq

\subsubsection {Contribution to the Lamb Shift}

Consider now contribution induced by the spanning photon  diagram with
"left" self-energy insertion in Fig.2e to the Lamb shift. We will below
repeat with minor changes considerations of the previous subsection.

General expression for the spanning photon diagram coincides with the one
in \eq{genspan}, only explicit expressions for the numerator structures
slightly change.

Unlike transformations in the previous sections we have to preserve
temporarily small nonvanishing virtuality $\rho=1-p^2$ of external electron
lines while combining denominators. This virtuality will be put to be equal
to zero in the final formulae but it is necessary to preserve it on
intermediate stages to qualify spurious infrared divergences which appear in
the subtraction term below and cancel one another. Hence, we use instead of
\eq{comb} slightly modified formulae

\beq             \label{combspanleft}
\Delta_l=t[{\bf k}^2u(1-ut)+t+\frac{x(1-u)}{1-x}+\rho]\equiv ta_l
\eeq

Explicit expression for the term ${\cal N}_F^{e}$ differs from the one used
in the previous subsection only due to the change of free Lorentz indices

\beq
{\cal N}_F^{e}= \gamma^\sigma(\hat p+\hat q)\gamma_0(\hat p+\hat q-\hat
k+1)\gamma_0(\hat p+\hat q+1)\gamma_\sigma,
\eeq

while we use another separation of different entries in the term ${\cal
N}_L^{e}$ here

\beq
{\cal N}_L^{e}={2}\hat q(\hat p+\hat q)\gamma_0(\hat p+\hat q-\hat
k+1)\gamma_0(\hat p +\hat q+1)\hat q
\eeq
\[
=(q^2+2pq)\{[2q^2\gamma_0(\hat p+\hat q-\hat k+1)\gamma_0+2\hat q\hat
p\gamma_0\hat q\gamma_0]-2\hat q\hat p\gamma_0\hat k\gamma_0\}
\]
\[
+q^22\hat q\hat p\gamma_0(\hat p+1)\gamma_0+4(pq)\hat q\hat p\gamma_0(\hat
p+1)\gamma_0
\]
\[
\equiv (q^2+2pq)[q^2{\cal N}_{L1}^{e}+{\cal N}_{L2}^{e}]+q^2{\cal N}_{L3}^{e}
+{\cal N}_{L4}^{e}.
\]

Repeating calculations performed in the previous section we obtain
(compare \eq{leftsigmsp})

\beq  \label{leftsigmspl}
\Xi_{00}^{e}=\frac{3}{8}(\frac{\alpha}{\pi})^2 \int_0^1dx\frac{1}{1-x}
\int_0^1du\int_0^1dt
\eeq
\[
\int\frac{d^4{q}}{i\pi^2}\{3(1-u)t^2\frac{{\cal N}_F^{e}(q-Q)
+{\cal N}_{L3}^{e}(q-Q)}
{(q^2-\Delta_l)^4}
\]
\[
+\frac{tx{\cal N}_{L1}^{e}(q-Q)}{(q^2-\Delta_l)^3}
+\frac{3t(1-t)x{\cal N}_{L2}^{e}(q-Q)}{(q^2-\Delta_l)^4}
\]
\[
+\frac{12t(1-u)t^2{\cal N}_{L4}^{e}(q-Q)}{(q^2-\Delta_l)^5}\}.
\]

Performing next integration over momentum we obtain (compare \eq{spanleft})
for the electron factor

\beq        \label{spleftgenl}
{\cal L}^{e}=\frac{3}{16}\int_0^1dx\frac{1}{1-x}\int_0^1du\int_0^1dt
\{\frac{t(1-u)(3t-5)+x(t-3)}{a_l}
\eeq
\[
+\frac{{\bf k}^2}{a_l^2}u[x(1-t)+(1-u)t^2(3ut-3u-2)
+\frac{1}{a_l^2}t(1-u)(t^2+t-6)
\]
\[
+2(1-u)[\frac{(2-t)}{a_l^2}-\frac{4t(1-t)}{a_l^3}]\}.
\]

Next it is necessary to perform subtraction of the part proportional to ${\bf
k}^2$ from this expression. Note that the separate terms in the last
brackets in \eq{spleftgenl} would lead to infrared divergent integrals at
${\bf k}^2=0$ if one omits small virtuality of the external electron line
introduced in the beginning of this subsection. However, all integrals are
perfectly convergent with nonvanising virtuality and we may use
auxiliary identity from \eq{refer} (for $n=2$ and $\eta=1$) to simplify the
integral representation. After this simplification integral representation
admits vanishing virtuality even at ${\bf k}^2=0$ and we obtain after
subtraction

\beq
\frac{L^{e}}{{\bf k}^4}           \label{spleftgens}
\equiv\frac{{\cal L}^{e}({\bf k})-{\cal L}^{e}(0)}{{\bf k}^2}
\eeq
\[
=\frac{3}{16}\int_0^1\frac{dx}{1-x}\int_0^1du
\int_0^1dt\{\frac{u(1-ut)}{a_la_{l0}}[x(3-t)+(1-u)t(5-3t)]
\]
\[
-t^2(1+t)u(1-u)(1-ut)(\frac{1}{a_l^2a_{l0}}+\frac{1}{a_la_{l0}^2})
+\frac{u}{a_l^2}[x(1-t)+(1-u)t^2(-2-3u+3ut)]
\]
\[
-\frac{8}{a_l^3}u^2(1-u)t(1-t)\}.
\]

Substituting subtracted electron factor in \eq{lamb} and integrating over
${\bf k}$ we obtain

\beq               \label{leftsplamb}
2\Delta E_{Lamb}^{e}=\:m\:\left(\frac{m_r}{m}\right)^3
\frac{\alpha^2(Z\alpha)^5}{\pi n^3}(\frac{3}{2\pi})
\int_0^1\frac{dx}{1-x}\int_0^1\frac{du}{\sqrt{1-u}}
\int_0^1\frac{dt}{\sqrt{1-ut}}
\eeq
\[
\{\frac{g^e_{3/2}}{a_{l0}^\frac{3}{2}}+\frac{g^e_{5/2}}{a_{l0}^\frac{5}{2}}\},
\]

where

\beq
g^{e}_{3/2}=xu[1-t+2(1-ut)(3-t)]+ut(1-u)[t(-2-3u+3ut)+2(1-ut)(5-3t)],
\eeq
\[
g^{e}_{5/2}=-3u(1-u)t[2u(1-t)+t(1-ut)(1+t)].
\]

Numerically we have

\beq
2\Delta E_{Lamb}^{e}=5.056(1)\:m\:\left(\frac{m_r}{m}\right)^3
\frac{\alpha^2(Z\alpha)^5}{\pi n^3}
\eeq

\section {Calculation of Contributions Induced by the Diagrams with
Insertions of the "Right" Self-Energy Operator}

\subsection {Diagrams Containing the Two-Loop Vertex}

\subsubsection {General Expression for the "Left" Vertex with Right
One-Loop Self-Energy Insertion}

Let us consider vertex diagram which is contained as a subgraph in Fig.2f.
Insertion of the one-loop electron self-energy in the FY gauge corresponds
to replacement (compare \eq{sesubl})

\beq        \label{sesubr}
\frac{1}{\hat p+\hat q-\hat k-1}\rightarrow \frac{3\alpha}{4\pi}(\hat
p+\hat q-\hat k)
\int_0^1dx\frac{x}{1-x}\frac{1}{q^2+2q(p-k)-\frac{x}{1-x}}
\eeq

Formal expression for the vertex has then the form

\beq                           \label{rightvert}
\Lambda_\mu^{f}
=3(\frac{\alpha}{4\pi})^2 \int_0^1dx\frac{x}{1-x}
\int\frac{d^4{q}}{i\pi^2}\frac{N_F^f+N_L^f/q^2}{q^2[q^2+2qp](q^2+2q(p-k)+k^2
-\frac{x}{1-x})},
\eeq

where

\beq
N_F^{f}= \gamma^\sigma(\hat p+\hat q+1)\gamma_\mu(\hat p+\hat q-\hat
k)\gamma_\sigma,
\eeq
\beq
N_L^{f}={2}\hat q(\hat p+\hat q+1)\gamma_\mu(\hat p+\hat q-\hat
k)\hat q.
\eeq

It is convenient to put down second numerator in the form

\beq
N_L^{f}=(q^2+2pq)[q^2N_{L1}^{f}+N_{L2}^{f}],
\eeq

where

\beq
N_{L1}^{f}=2\gamma_\mu,
\eeq
\[
N_{L2}^{f}=2\gamma_\mu(\hat p-\hat k)\hat q.
\]

Combining denominators in eq.(\ref{rightvert}) (compare \eq{comb})

\beq      \label{combr}
(1-t)q^2+t[(1-u)(q^2+2qp)
\eeq
\[
+u(q^2+2q(p-k)-\frac{x}{1-x})]=(q+Q)^2-\Delta_r,
\]
\[
Q=pt-kut,
\]
\[
\Delta_r=t[{\bf k}^2u(1-ut)+t+\frac{xu}{1-x}]\equiv ta_r,
\]

we obtain representation for the bare vertex operator in the form

\beq
\Lambda_\mu^{f}=3(\frac{\alpha}{4\pi})^2 \int_0^1dx\frac{x}{1-x}
\int_0^1du\int_0^1dt\int\frac{d^4{q}}{i\pi^2}\{2t\frac{N_F^{f}(q-Q)}
{(q^2-\Delta_r)^3}
\eeq
\[
+\frac{N_{L1}^{f}(q-Q)}{(q^2-\Delta_{r1})^2}
+\frac{2(1-t)N_{L2}^{f}(q-Q)}{(q^2-\Delta_{r1})^3}\},
\]

where

\beq
\Delta_{r1}=\Delta_r(u=1)=t[{\bf k}^2(1-t)+t+\frac{x}{1-x}+\rho]\equiv
ta_{r1}.
\eeq

Next we perform shift of integration variable over momentum and euclidean
rotation (and we slightly change notation below as $q$ further means
euclidean shifted momentum) and obtain

\beq
\Lambda_\mu^{f}=3(\frac{\alpha}{4\pi})^2 \int_0^1dx\frac{x}{1-x}
\int_0^1du\int_0^1dt\int_0^\infty{dq^2}q^2\{\frac{2t}{(q^2+\Delta_r)^3}
\eeq
\[
[[3q^2 -2{\bf k}^2ut(1-ut)-2(1-t^2)]\gamma_\mu
-2t(1-t)\gamma_\mu(\hat p-\hat k-1)
\]
\[
+4(1-t)(1-ut)p_\mu(\hat p-\hat k-1)
+2(1-t)^2\gamma_\mu\hat k]
\]
\[
+\frac{2}{(q^2+\Delta_{r1})^2}\gamma_\mu
-\frac{4t(1-t)({\bf k}^2-1)}{(q^2+\Delta_{r1})^3}\gamma_\mu\}.
\]

Momentum integration leads to

\beq                                 \label{unsubtrr}
\Lambda_\mu^{f}=3(\frac{\alpha}{4\pi})^2 \int_0^1dx\frac{x}{1-x}
\int_0^1du\int_0^1dt\{[2t(\log{\frac{\Lambda^2+\Delta_r}{\Delta_r}}
-\frac{3}{2}\frac{\Lambda^2}{\Lambda^2+\Delta_r})
\eeq
\[
-\frac{2{\bf k}^2ut(1-ut)+2(1-t^2)}{a_r}]\gamma_\mu
-\frac{2t(1-t)}{a_r}\gamma_\mu(\hat p-\hat k-1)
\]
\[
+\frac{4(1-t)(1-ut)}{a_r}p_\mu(\hat p-\hat k-1)
+\frac{2(1-t)^2}{a_r}\gamma_\mu\hat k
\]
\[
+2[\log{\frac{\Lambda^2+\Delta_{r1}}{\Delta_{r1}}}-\frac{\Lambda^2}{\Lambda^2
+\Delta_{r1}}]\gamma_\mu+\frac{2(1-t)(1-{\bf k}^2)}{a_{r1}}\gamma_\mu\}.
\]

Subtraction term is equal to

\beq          \label{subtrr}
\Lambda_\mu^{f}(0)=3(\frac{\alpha}{4\pi})^2
\int_0^1dx\frac{x}{1-x}
\int_0^1du\int_0^1dt\{2t(\log{\frac{\Lambda^2+\Delta_{r0}}{\Delta_{r0}}}
-\frac{3}{2}\frac{\Lambda^2}{\Lambda^2+\Delta_{r0}})
\eeq
\[
-\frac{2(1-t^2)}{a_{r0}}
+2[\log{\frac{\Lambda^2+\Delta_{r10}}{\Delta_{r10}}}-\frac{\Lambda^2}{\Lambda^2
+\Delta_{r10}}]+\frac{2(1-t)}{a_{r10}}\}
\gamma_\mu,
\]

where

\[
a_{r0}={a_r}_{|{\bf k}=0},
\]
\[
a_{r10}={a_r}_{1{|{\bf k}=0}}.
\]

We extract explicit dependence on ${\bf k}^2$ in \eq{unsubtrr} after
subtraction with the help of integration by parts

\beq
\int_0^1du\log{\frac{a_{r0}}{a_r}}={\bf
k}^2\int_0^1du\frac{t(1-u)}{a_r}(u-\frac{1-ut}{a_{r0}}),
\eeq
\[
\int_0^1dt\log{\frac{a_{r10}}{a_{r1}}}=-{\bf
k}^2\int_0^1\frac{dt}{a_{r1}}\frac{t}{x+t(1-x)}.
\]

Then renormalized vertex with the right self-energy insertion acquires the
form (compare \eq{leftverrgtse})

\beq         \label{rightvertren}
\Lambda_\mu^{f}=3(\frac{\alpha}{4\pi})^2 \int_0^1dx\frac{x}{1-x}
\int_0^1du\int_0^1dt\{{\bf k}^2{\cal A}_r\gamma_\mu
\eeq
\[
+{\cal B}_r\gamma_\mu(\hat p-\hat k-1)
+{\cal C}_rp_\mu(\hat p-\hat k-1)
+{\cal E}_r\gamma_\mu\hat k\},
\]

where

\beq
{\cal A}_r=-\frac{2ut(1-t)}{a_r}+\frac{2(1-ut)(u-t^2)}{a_ra_{r0}}
-\frac{2}{a_{r1}[x+t(1-x)]},
\eeq
\[
{\cal B}_r=-\frac{2t(1-t)}{a_r},
\]
\[
{\cal C}_r=\frac{4(1-t)(1-ut)}{a_r},
\]
\[
{\cal E}_r=\frac{2(1-t)^2}{a_r}.
\]

Note that as well as in the case of the left self-energy insertion in \eq{aa}
and \eq{bcd} all four functions ${\cal A}_r({\bf k})$, ${\cal B}_r({\bf
k})$, ${\cal C}_r({\bf k})$ and ${\cal E}_r({\bf k})$ are finite at ${\bf
k}=0$.

\subsubsection{Contribution to HFS Interval}

It is not difficult now to obtain contribution to HFS

\beq               \label{righthfs}
2\Delta E_{HFS}^{f}=\frac{\alpha^2(Z\alpha)}{\pi
n^3}E_F(\frac{3}{\pi^2})\int_0^1dx\frac{x}{1-x}
\int_0^1du\int_0^1dt\int_0^\infty d|{\bf k}|[{\cal A}_r({\bf k})
\eeq
\[
-\frac{{\cal E}_r({\bf k})-{\cal E}_r(0)}{{\bf k}^2}],
\]

where

\beq
{\cal A}_r({\bf k})-\frac{{\cal E}_r({\bf k})-{\cal E}_r(0)}{{\bf k}^2}
=2\{-\frac{ut(1-ut)}{a_r}+\frac{(1-ut)[2u(1-t)-(1-u)t^2]}{a_ra_{r0}}
\eeq
\[
-\frac{1}{a_{r1}(x+t-xt)}\}.
\]
Subtraction of ${\cal E}_r(0)$ in eq.(\ref{righthfs}) corresponds to
subtraction of the contribution to HFS induced by the second order
anomalous magnetic moment which is of lower order in parameter $Z\alpha$.

After momentum integration we obtain three-dimensional integral for the
contribution to HFS induced by both diagrams with nonsymmetric self-energy
insertions in Fig.2f

\beq
2\Delta E_{HFS}^{f}=\frac{\alpha^2(Z\alpha)}{\pi
n^3}E_F\{-3+\frac{3}{\pi}\int_0^1dx\frac{x}{1-x}
\int_0^1\frac{du}{\sqrt{u}}\int_0^1\frac{dt}{\sqrt{1-ut}}
\eeq
\[
[\frac{f^{f}_{1/2}}{a_{r0}^\frac{1}{2}}+\frac{f^{f}_{3/2}}
{a_{r0}^\frac{3}{2}}\},
\]

where

\beq
f^{f}_{1/2}=-ut(1-t),
\eeq
\[
f^{f}_{3/2}=(1-ut)[2u(1-t)-(1-u)t^2].
\]

Numerically we obtain

\beq
2\Delta E_{HFS}^{f}=-3.401(1)\frac{\alpha^2(Z\alpha)}{\pi n^3}E_F.
\eeq

\subsubsection{Contribution to the Lamb Shift}

Contribution to the Lamb shift induced by the diagram in
Fig.2f has the same form as in \eq{leftlmb}, where

\beq
\frac{{\cal L}^{f}({\bf k})-{\cal L}^{f}(0)}{{\bf
k}^2}=\frac{u(1-t)(1-ut)}{a_ra_{r0}}
\eeq
\[
-\frac{2u(1-ut)^2(u-t^2)}{a_ra_{r0}^2}+\frac{2(1-x)(1-t)}{a_{r1}[x+t(1-x)]^2}.
\]

After momentum integration we obtain respective contribution to the Lamb
shift in the form of the three-dimensional integral

\beq
2\Delta E_{L}^{f}=m\:(\frac{m_r}{m})^3
\frac{\alpha^2(Z\alpha)^5}{\pi
n^3}\{-2+\frac{3}{\pi}\int_0^1dx\frac{x}{1-x}
\int_0^1du\int_0^1dt
\eeq
\[
\sqrt{t(1-ut)}[\frac{g^{f}_{3/2}}{a_{r0}^\frac{3}{2}}
+\frac{g^{f}_{5/2}}{a_{r0}^\frac{5}{2}}]\},
\]

where

\beq
g^{f}_{3/2}=-1+t,
\eeq
\[
g^{f}_{5/2}=2(1-ut)(u-t^2).
\]

Numerically we obtain

\beq
2\Delta E_{L}^{f}=-1.017(1)m\:(\frac{m_r}{m})^3
\frac{\alpha^2(Z\alpha)^5}{\pi n^3}.
\eeq

\subsection {Spanning Photon Diagram with Symmetrical
Self-Energy Insertion}

\subsubsection {Contribution to HFS Interval}

Consider now spanning photon  diagram with symmetrical self-energy insertion
in Fig.2g. Unlike discussion of the vertex in the previous subsection we
will consider here only the part of the spanning photon diagram relevant
for hyperfine splitting. Insertion of the one-loop electron self-energy in
the FY gauge corresponds to replacement described in eq.(\ref{sesubr}) in
the central propagator in the spanning photon diagram.

Formal expression for the spanning photon diagram has the form

\beq
\Xi_{\mu\nu}^{g}         \label{genspansym}
=3(\frac{\alpha}{4\pi})^2 \int_0^1dx\frac{x}{1-x}
\eeq
\[
\int\frac{d^4{q}}{i\pi^2}\frac{N_F^g+N_L^g/q^2}{q^2(q^2+2qp-\rho)^2
[q^2+2q(p-k)+k^2-\frac{x}{1-x}-\rho]},
\]

where

\beq
N_F^{g}=\gamma^\sigma(\hat p+\hat q+1)\gamma_\mu(\hat p+\hat q-\hat
k)\gamma_\nu(\hat p+\hat q+1)\gamma_\sigma,
\eeq

\beq
N_L^{g}={2}\hat q(\hat p+\hat q+1)\gamma_\mu(\hat p+\hat q-\hat
k)\gamma_\nu(\hat p +\hat q+1)\hat q
\eeq
\[
={2(q^2+2pq)^2}\gamma_\mu(\hat q-\hat p)\gamma_\nu.
\]

Note that we temporarily preserve in \eq{genspansym} nonvanishing virtuality
$\rho=1-p^2$ of the external fermion lines. This virtuality will be put to be
equal to zero in
the final formulae but it is necessary to preserve it on intermediate stages
to qualify spurious infrared divergences which cancel one another. The
problem of infrared divergences is even more acute here than in discussion
of the Lamb shift contribution induced by the diagram in Fig. 2e, since
integral in \eq{genspansym} contains more powers of integration momentum $q$
in the denominator. We will see below that as a result separate parametric
integrals giving contributions to the electron factor diverge (for vanishing
virtuality) even in the case of nonvanishing exchanged momentum ${\bf k}$
and only total contribution to the electron factor turns out to be finite.

Envisaging problems with infrared divergences it is convenient prior to
integration to separate in the numerators $N_F^{g}$ and $N_L^{g}$ infrared
safe terms  $N_F''^{g}$ and $N_{L2}^{g}$

\beq          \label{numirf}
N_F^{g}=N_{F1}^{g}+\gamma^\sigma(\hat p+1)\gamma_\mu(\hat p+\hat q-\hat
k)\gamma_\nu(\hat p+1)\gamma_\sigma
\eeq
\[
=N_{F1}^{g}+4\gamma_\mu(\hat q-\hat k)\gamma_\nu\equiv N_{F1}^{g}+N_{F2}^{g},
\]

and

\beq                           \label{numirl}
N_L^{g}=q^2N_{L1}^{g}+{8(pq)^2}\gamma_\mu(\hat q-\hat k)\gamma_\nu
\eeq
\[
\equiv q^2N_{L1}^{g}+N_{L2}^{g},
\]

Next we combine denominators as in eq.(\ref{combr}) but preserving
nonvanishing virtuality of external electron lines so that explicit
expression for $\Delta_r$ slightly changes and below

\beq
\Delta_r=t[{\bf k}^2u(1-ut)+t+\frac{xu}{1-x}+\rho]\equiv ta_r.
\eeq

After shift of integration variable $q\rightarrow q-Q$ (see \eq{combr}) we
obtain retaining only those numerator structures which are relevant for
contribution to HFS

\beq
\Xi_{\mu\nu}^{g}               \label{spsymgen}
=\frac{3}{16}(\frac{\alpha}{\pi})^2 \int_0^1dx\frac{x}{1-x}\int_0^1du2(1-u)
\int_0^1dt3t^2
\eeq
\[
\int\frac{d^4{q}}{i\pi^2}\{\frac{N_{F2}^{g}(q-Q)
+[N_{F1}^{g}(q-Q)+N_{L1}^{g}(q-Q)]}{(q^2-\Delta_r)^4}
\]
\[
+4(1-t)\frac{N_{L2}^{g}(q-Q)}{(q^2-\Delta_r)^5}\}.
\]

After integration over momentum we obtain

\beq     \label{finspansym}
F^{g}
=\frac{3}{4}\int_0^1dx\frac{x}{1-x}\int_0^1
du(1-u)\int_0^1dt
\{(1-2ut)[\frac{2-t}{a_r^2}-\frac{4t(1-t)}{a_r^3}]
\eeq
\[
+\frac{3t(1-2ut)}{a_r}+\frac{2t(1-ut)}{a_r^2}(3-t+{\bf
k}^2u^2t)\}.
\]

infrared unsafe terms in the central brackets in \eq{finspansym} may be
easily exorcised with  the help of the identity in\eq{refer}. Substituting
then expression electron factor in \eq{finspansym} in \eq{hfs} and
integrating over ${\bf k}$ one obtains contribution to HFS interval

\beq               \label{spsymhfs}
\Delta E_{HFS}^{g}
=\frac{\alpha^2(Z\alpha)}{\pi n^3}E_F
(\frac{3}{2\pi})\int_0^1dx\frac{x}{{1-x}}\int_0^1
\frac{du}{\sqrt{u}}(1-u)\int_0^1dt\frac{t}{\sqrt{1-ut}}
\{\frac{f^{g}_{1/2}}{\sqrt{a_{r0}}}
\eeq
\[
+\frac{f^{g}_{3/2}}{a_{r0}^\frac{3}{2}}\},
\]

where

\beq
f^{g}_{1/2}=3-5ut
\eeq

and

\[
f^{g}_{3/2}=-u(1-t)-t(1-ut).
\]

Numerically we obtain

\beq
\Delta E_{HFS}^{g}=2.682(1)\frac{\alpha^2(Z\alpha)}{\pi
n^3}E_F.
\eeq

\subsubsection {Contribution to the Lamb Shift}

Consider now contribution induced by the spanning photon  diagram with
symmetrical self-energy insertion in Fig.2g to the Lamb shift. We will
closely follow considerations performed in the previous subsection.

Formal expression for the spanning photon diagram has the same form
as in \eq{genspansym}, the only difference is connected with the free
Lorentz indices.

We separate infrared safe terms in the numerators in the same way as in the
previous subsection (see \eq{numirf} and \eq{numirl}) and obtain  repeating
all by now standard steps

\beq     \label{nfinspansymlmb}
{\cal L}^{g}
=\frac{3}{8}\int_0^1dx\frac{x}{1-x}\int_0^1
du(1-u)\int_0^1dtt
\{\frac{3(2t -1)}{2a_r}
\eeq
\[
+\frac{{\bf k}^2u(2t^2u - t - 1)+ 2(3t-2)}{a_r^2}
+\frac{4(1-t)(t-{\bf k}^2u^2)}{a_r^3}\}.
\]

This expression may be further simplified with the help of identity similar
to the one in \eq{refer} but obtained with the help of integration by parts
of the fraction $t^2(1-t)/a_r$ instead of the fraction in \eq{apintp}. We
then obtain

\beq     \label{finspansymlmb}
{\cal L}^{g}
=\frac{3}{8}\int_0^1dx\frac{x}{1-x}\int_0^1
du(1-u)\int_0^1dtt
\{\frac{3(2t -1)}{2a_r}
\eeq
\[
+\frac{{\bf k}^2u(2t^2u - t - 1)}{a_r^2}
-\frac{4u^2(1-t)^2{\bf k}^2}{a_r^3}\}.
\]

This expression is very convenient for subtraction of the low frequency
asymptote since almost all terms in it contain factor ${\bf k}^2$
explicitly. Performing subtraction we obtain

\beq
\frac{L^{g}}{{\bf k}^4}=\frac{{\cal L}^{g}({\bf k})-{\cal L}^{g}(0)}{{\bf
k}^2}=\frac{3}{8}\int_0^1dx\frac{x}{1-x}\int_0^1 du(1-u)\int_0^1dtt
\eeq
\[
\{-\frac{3u(1-ut)(2t -1)}{2aa_{r0}}
+\frac{u(2t^2u - t - 1)}{a_r^2}
-\frac{4u^2(1-t)^2}{a_r^3}\}.
\]

Substituting now subtracted electron factor in \eq{lamb} and performing
integration over ${\bf k}$ we obtain contribution to the Lamb shift

\beq               \label{spsymlamb}
\Delta E_{Lamb}^{g}=\:m\:\left(\frac{m_r}{m}\right)^3
\frac{\alpha^2(Z\alpha)^5}{\pi n^3}
(\frac{3}{\pi})\int_0^1dx\frac{x}{1-x}\int_0^1
du\sqrt{u}(1-u)\int_0^1dt\frac{t}{\sqrt{1-ut}}
\eeq
\[
\{\frac{g^{g}_{3/2}}{a_{r0}^\frac{3}{2}}+\frac{g^{g}_{5/2}}
{a_{r0}^\frac{5}{2}}\},
\]

where

\beq
g^{g}_{3/2}=2-7t-3ut+8ut^2,
\eeq
\[
g^{g}_{5/2}=-3u(1-t)^2.
\]

Numerically we obtain

\beq
\Delta E_{Lamb}^{g}=-0.14601(4)\:m\:\left(\frac{m_r}{m}\right)^3
\frac{\alpha^2(Z\alpha)^5}{\pi n^3}.
\eeq

\section {Calculation of Contribution Induced by Insertion of the Rainbow
Self-Energy Operator}

\subsection{General Expression for the Rainbow Self-Energy Diagram}

Let us construct renormalized two-loop rainbow contribution to the electron
self-energy operator in the FY gauge. To this end we take general
expression for the one-loop bare self-energy operator in the FY
gauge\footnote{We use dimensional momenta in this section}

\beq
\Sigma(p)=\frac{\alpha}{4\pi}\int\frac{d^4q}{i\pi^2q^2}(g^{\alpha\beta}
+2\frac{q^\alpha q^\beta}{q^2})\gamma_\alpha\frac{1}{\hat p+\hat
q-m}\gamma_\beta
\eeq

and perform substitution similar to the one in \eq{sesubr}, which describes
insertion of the renormalized self-energy operator in the FY gauge on the
internal line.  We then obtain

\beq
\Sigma_B^R(p)=3(\frac{\alpha}{4\pi})^2\int_0^1dx\frac{x}{1-x}
\int\frac{d^4q}{i\pi^2}\frac{N_F^{h}+N_L^{h}/q^2}
{q^2[(p+q)^2-\frac{m^2}{1-x}]},
\eeq

where

\beq
N_F^{h}=\gamma^\sigma(\hat p +\hat q)\gamma_\sigma=-2(\hat p+\hat q),
\eeq
\[
N_L^{h}=2\hat q(\hat p+\hat q)\hat q=q^2(2\hat q-\hat p)+(2\hat q\hat p\hat
q+q^2\hat p)\equiv q^2N_{L1}^{h} +N_{L2}^{h}.
\]

Combining denominators with the help of the Feynman parameters and
performing shift of integration momentum\footnote{Note that $N_F^{h}
+N_{L1}^{h}=-3\hat p$ does not contain linear in the integration momentum term,
while integral
depending on $N_{L2}^{h}$ is UV finite, and, hence, no surface terms appear
under shift of the integration momentum.}, Wick rotation and momentum
integration we obtain explicit expression for the unrenormalized rainbow
self-energy operator

\beq
\Sigma_B^R(p)=3(\frac{\alpha}{4\pi})^2 \int_0^1dx\frac{x}{1-x}\int_0^1dt
\left[-3\hat pH(p)-\frac{3\hat
pp^2t^2(1-t)}{\Delta}\right],
\eeq

where

\beq          \label{uvself}
H(p)=\log{\frac{\Lambda^2+\Delta_h}{\Delta_h}}-
\frac{\Lambda^2}{\Lambda^2+\Delta_h},
\eeq
\[
\Delta_h=t[-p^2(1-t)+\frac{m^2}{1-x}].
\]

Term $\Delta_h$ is preserved on the background of the UV cutoff in
\eq{uvself} due to the same reasons which were explained after \eq{subtr}.
It is convenient to rewrite $H(p)$ in the form

\beq
H(p)=H(m)+[H(p)-H(m)],
\eeq

where first term is the momentum independent constant which in any case
disappears after subtraction and the second term already admits limit of the
infinite cutoff. We integrate the this second term over $t$ by parts to get
rid of logarithm and obtain

\beq
\Sigma_B^R(p)=-9(\frac{\alpha}{4\pi})^2 \hat
p\int_0^1dx\frac{x}{1-x}\int_0^1dt
\left[H(m)+\frac{p^2t^2(2-t)}{\Delta_h}
\right].
\eeq

After two standard subtractions on the mass-shell followed by change of
variables $v=1-t$, $\xi=v(1-x)$ and integration by parts over new variable
$v$ we come to the representation for the FY gauge renormalized rainbow
mass-operator which is the most convenient one for subsequent calculation
of the contributions to the energy splitting

\beq   \label{sexi}
\Sigma^R(p)=-9(\frac{\alpha}{4\pi})^2(\hat p-m)^2
\int_0^1\frac{dv}{v}(1-v)^2\int_0^v\frac{d\xi}{(1-\xi)^2}\frac{\hat p
(1+\xi)+2m}{m^2-p^2\xi}.
\eeq

Additional integration by parts over $v$ leads to the representation in the
form of one-dimensional integral

\beq               \label{sev}
\Sigma^R(p)=9(\frac{\alpha}{4\pi})^2(\hat p-m)^2
\int_0^1dv\left[\frac{1-v+\log
v}{(1-v)^2}+\frac{1}{2}\right]\frac{\hat p(1+v)+2m}{m^2-p^2v}
\eeq
\[
\equiv(\hat p-m)^2(\hat p\sigma_p+m\sigma_m).
\]

Even remaining final integration may be performed analytically and one can
obtain representation of the rainbow contribution to the electron
self-energy in the closed form

\beq
\Sigma^R(p)=\frac{3\alpha}{8\pi}\Sigma^{FY}(p)
+\frac{9}{16}(\frac{\alpha}{\pi})^2\frac{(\hat p-m)^2}{m^2}\{2(\hat
p+m){\cal F}_1(\rho)
\eeq
\[
-(\hat p+2m){\cal F}_2(\rho)-m\frac{\log{\rho}}{1-\rho}\},
\]

where $\rho$ is the virtuality of the electron line (momentum $p$ is
arbitrary in this subsection)

\beq
\rho=\frac{m^2-p^2}{m^2},
\eeq

$\Sigma^{FY}(p)$ is the one-loop electron self-energy (compare \eq{sigma})

\[
\Sigma^{FY}(p)=-\frac{3\alpha\hat p}{4\pi}(\hat
p-m)^2[\frac{1}{1-\rho}+\frac{\rho}{(1-\rho)^2}\log{\rho}]
\]

and functions ${\cal F}_i$ are defined as follows\footnote{We use
standard Euler dilogarithm Li$_2(\rho)$ here unlike function Li$(\rho)$
in \eq{phi} which was defined as the real part of the Euler dilogarithm.}

\beq
{\cal F}_1(\rho)=\frac{1}{\rho^2}[\frac{\pi^2}{6}-\rm{Li}_2(1-\rho)
+\rho(\log{\rho}-1)],
\eeq
\[
{\cal F}_2(\rho)=\frac{1}{\rho}[\frac{\pi^2}{6}-\rm{Li}_2(1-\rho)].
\]

\subsubsection {Contribution to HFS Interval}

For final calculation of the contribution to the HFS interval we use
representation (\ref{sexi})

\beq
\Delta E_{HFS}^{h}=\frac{\alpha^2(Z\alpha)}{\pi
n^3}E_F(\frac{9}{2\pi^2})\:\int_0^\infty dk\int_0^1\frac{dv}{v}(1-v)^2
\int_0^v\frac{d\xi}{(1-\xi)^2}\frac{1+\xi}{1-\xi+{\bf k}^2\xi}
\eeq
\[
=\frac{\alpha^2(Z\alpha)}{\pi
n^3}E_F(\frac{9}{4\pi})\int_0^1\frac{dv}{v}(1-v)^2
\int_0^v\frac{d\xi}{\xi^{\frac{1}{2}}}\frac{1+\xi}{(1-\xi)^\frac{5}{2}}
=\frac{33}{16}\frac{\alpha^2(Z\alpha)}{\pi n^3}E_F.
\]

\subsubsection {Contribution to the Lamb Shift}

With the help of representation (\ref{sexi}) we obtain contribution to the
Lamb shift interval
\beq
\Delta E_{L}^{h}=
\:m\:\left(\frac{m_r}{m}\right)^3
\frac{16(Z\alpha)^5}{\pi n^3}
\int_0^\infty\frac{dk}{k^2}\{[\sigma_p(k)+\sigma_m(k)]
-[\sigma_p(0)+\sigma_m(0)]\}
\eeq
\[
=\:m\:\left(\frac{m_r}{m}\right)^3
\frac{\alpha^2(Z\alpha)^5}{\pi n^3}\left(-\frac{9}{\pi^2}\right)
\int_0^1\frac{dv}{v}(1-v)^2
\int_0^v\frac{d\xi}{(1-\xi)^2}[\frac{3+\xi}{1-\xi+{\bf k}^2\xi}
-\frac{3+\xi}{1-\xi}]
\]
\[
=\frac{153}{40}\:m\:\left(\frac{m_r}{m}\right)^3
\frac{\alpha^2(Z\alpha)^5}{\pi n^3}.
\]

\section{Discussion of Results}

We presented above results of calculation of contributions to the Lamb shift
and HFS of order $\alpha^2(Z\alpha)^5$ induced by all two-loop insertions in
the electron line containing as a block one-loop electron self-energy graph.
Calculations were performed in the FY gauge which is the most suitable one
due to its infrared smoothness. The formulae for different contributions
obtained above admit numerical calculation with arbitrary accuracy. We
consider it a bit premature to present here the sum of all contributions
obtained above since considered set of graphs is not gauge invariant and
comparison with the experimental data have to await until all other
contributions would be obtained. Nevertheless we have chosen to present
above calculations with sufficient details both because of their volume and
to present main technical tricks used in our work.

Results of calculation of other contributions will be published separately.

\medskip

This paper was completed during the visit of one of the authors (M.E.) to
the Penn State University. M.E. is deeply grateful to his colleagues at Penn
State and especially to Prof. H. Grotch for kind hospitality.

\medskip

This work was supported by the Russian Foundation for Fundamental Research
under grant \#93-02-3853. Work by S.G. Karshenboim was also supported by a
Soros Foundation grant awarded by the American Physical Society.

\appendix
\section{}

Calculations performed in the main body of this paper were greatly
facilitated by the infrared smoothness of the FY gauge. We would
like to display in this appendix how cancellation of infrared divergences in
the FY gauge occurs and derive a useful identity widely used above.

Typical integral for the contribution to the electron factor with the
worst infrared behavior has the form

\beq
I=\int\frac{d^4q}{\pi^2i}\frac{1}{q^2}(g^{\alpha\beta}+\xi\frac{q^\alpha
q^\beta}{q^2})
\eeq
\[
\int_0^1dx\ldots\int_0^1dzf(x,\ldots,z)
\frac{\gamma_\alpha(\hat p+m)\hat A(\hat
p+m)\gamma_\beta}{[q^2+2q(p\eta-k\tau)-\omega]^{n+1}},
\]

where $p$ is the on-shell electron momentum $\hat p=m$ and matrix $\hat A$
and function of the Feynman parameters $\omega$ do not depend on $q$.

Taking into account mass-shell condition for the vector $p$ one easily
obtains

\beq
I=4\hat A\int_0^1dx\ldots\int_0^1dzf(x,\ldots,z)\int\frac{d^4q}{\pi^2i}
\{\frac{p^2}{q^2[q^2+2q(p\eta-k\tau)-\omega]^{n+1}}
\eeq
\[
+\frac{\xi(pq)^2}{q^4[q^2+2q(p\eta-k\tau)-\omega]^{n+1}}\}.
\]

We omit below unessential for further considerations integration over the
Feynman parameters and obtain after combining denominators, shift of
integration variable $q\rightarrow q-(p\eta-k\tau)t$ and the Wick rotation

\beq
I=(n+1)(-1)^{n+2}p^2\int_0^1dtt^n\int\frac{d^4q}{\pi^2}
\{\frac{1}{(q^2+\Delta)^{n+2}}
\eeq
\[
+\xi(n+2)(1-t)\frac{q^2/4-p^2\eta^2t^2-k^2\tau^2t^2}{(q^2+\Delta)^{n+3}}\},
\]

where

\beq
\Delta=(p^2\eta^2t+k^2\tau^2t+\omega)t\equiv ta.
\eeq

Integrating over $q$ one obtains

\beq         \label{exp}
I=\frac{(-1)^n}{n}m^2\int_0^1dt\{\frac{1+(\xi/2)(1-t)}{a^n}
-n\xi\frac{t(1-t)(m^2\eta^2+k^2\tau^2)}{a^{n+1}}\}.
\eeq

Integration over $t$ is in the general case infrared unsafe and may lead to
infrared divergences if, e.g. variable $\omega$ vanishes for some reason.
Hence, it would be helpful to gain additional power of $t$ in the numerator
of the integrand. To this end it is useful to observe validity of the
simple identity

\beq       \label{apintp}
\frac{\partial}{\partial
t}\frac{t(1-t)}{a^n}=\frac{1-2t}{a^n}-n\frac{t(1-t)}{a^{n+1}}
\frac{\partial a}{\partial t}\equiv
\frac{1-2t}{a^n}-n\frac{t(1-t)(m^2\eta^2+k^2\tau^2)}{a^{n+1}}.
\eeq

Integrating the last term in \eq{exp} with the help of the identity in
\eq{apintp} one obtains

\beq
I=\frac{(-1)^n}{n}m^2\int_0^1\frac{dt}{a^n}[1-\frac{\xi}{2}+\frac{3}{2}\xi
t].
\eeq

We see that terms with minimal power of $t$ in the numerator disappear when
$\xi=2$, i.e. in the FY gauge, thus making integration over $t$
more smooth for low $t$. We used this trick with integration by parts
abundantly in this paper and we put it down here for references in the
FY gauge (i.e. when $\xi=2$)

\beq           \label{refer}
\int_0^1dt\{\frac{2-t}{a^n}-2n\frac{t(1-t)(m^2\eta^2+k^2\tau^2)}{a^{n+1}}\}
=3\int_0^1dt\frac{t}{a^n}.
\eeq

Note that identity in \eq{refer} is only a representative of a large family
of identities which may be obtained in the same way but choosing different
numerators in \eq{apintp}.

\newpage

\begin{center}
\large{Figure Captions}
\end{center}

\noindent Fig.1. Six gauge-invariant sets of diagrams producing
contributions of order $\alpha^2(Z\alpha)^5$ to HFS and Lamb shift.

\medskip
\noindent
Fig.2. All graphs with two radiative photons producing contributions of
order $\alpha^2(Z\alpha)^5$ to HFS and Lamb shift.

\end{document}